**ARTICLE TITLE**

# Artificial Intelligence for early detection of circulatory shock in ICU patients

Jaume Aguiló Piña[1,2], Laia Subirats[1], Aina Frau-Pascual[2], Alba Gorriz[1,2], Rudys Magrans Nicieza[2] , Jordi Morillas Perez[3]

[1] NeuroADaS Lab, Universitat Oberta de Catalunya, Rambla del Poblenou 156, Barcelona, Spain, jaguilopi@uoc.edu, lsubirats@uoc.edu, albagorriz@gmail.com

[2] Better Care SL, C/ Pare Sallarès 153, Sabadell, Spain, jaguilo@bettercarehealth.com, afrau@bettercarehealth.com, albagorriz@gmail.com, rmagrans@bettercarehealth.com

[3] SCIAS-Hospital de Barcelona, Avinguda Diagonal 660, Barcelona, Spain, 31928jmp@gmail.com

Corresponding author:

Email: jaguilopi@uoc.edu

## ABSTRACT

Circulatory shock is one of the leading causes of mortality in intensive care units (ICUs), and its early detection is critical to enable timely treatment and improve clinical outcomes.

This study aimed to develop and evaluate a two-stage cascade machine learning framework for the early detection and etiological classification of circulatory shock in critically ill patients. Using data from the Medical Information Mart for Intensive Care (MIMIC)-IV database, four patient groups were defined: septic shock, cardiogenic shock, hypovolemic shock, and a non-shock control group, comprising a total of 32,907 patients. Vital signs and laboratory data were collected during the first six hours after ICU admission. After data cleaning and missing-value imputation, the mean value of each variable was used for model development.

Several machine learning algorithms were compared, including logistic regression, Random Forest, XGBoost, and multilayer perceptron (MLP) networks. Random Forest and XGBoost achieved the highest overall performance, with an AUROC of approximately 0.82-0.83 for shock detection, and a macro-averaged sensitivity of approximately 0.61 and precision of approximately 0.58 across all four classes. Classification performance was highest for the non-shock group, followed by septic and cardiogenic shock, while hypovolemic shock showed the lowest performance.

These results indicate that machine learning models can identify early signs of hemodynamic deterioration associated with circulatory shock and may support clinical decision-making in the ICU. However, further improvements are needed for the classification of specific shock subtypes, particularly hypovolemic and cardiogenic shock, as well as for real-time clinical implementation.

## 1. INTRODUCTION

The use of machine learning is expanding in healthcare, particularly in the early identification of various critical situations that can seriously compromise a patient's condition [1–7]. This is especially relevant for cases in which late detection can be fatal, and which, once established, are difficult for healthcare professionals to manage. The clearest example of this is circulatory shock, a complication that occurs in up to a third of patients admitted to the ICU [8,9]

Circulatory shock is defined as a physiological state of acute circulatory insufficiency in which the oxygen supply to the tissues is inadequate to meet their metabolic needs. This leads to tissue hypoperfusion, which, if not corrected early, triggers progressive dysfunction of vital organs and ultimately potentially irreversible multiple organ failure [8,10].

The three main and most frequent types of circulatory shock are:

- Hypovolemic shock, caused by a significant reduction in intravascular volume, which leads to decreased venous return, reduced cardiac output, and inadequate tissue perfusion. Generally as a result of acute hemorrhage, such as trauma or internal bleeding, although it can also result from severe fluid losses due to other conditions such as extensive burns [9,10].
- Cardiogenic shock, caused by an alteration in the contractile function of the heart, which results in decreased cardiac output despite adequate intravascular volume. The most common cause is acute myocardial infarction, but it may also arise from severe heart failure, arrhythmias, or mechanical complications affecting cardiac function [9,10].
- Septic shock, the most frequent in the ICU, is a severe manifestation of sepsis caused by a dysregulated host response to infection, which leads to generalised vasodilation, increased vascular permeability, ineffective blood flow distribution, often accompanied by profound hypotension [9,10].

Although other types of circulatory shock are recognised, such as obstructive, anaphylactic or neurogenic shock, their low prevalence and different pathophysiology mean that they are outside the scope of this article. Regardless of its etiology, circulatory shock is associated with high morbidity and mortality, as well as a substantial economic burden on healthcare systems. Early detection and accurate classification are therefore essential for improving patient outcomes. However, the current clinical approach relies on a complex and time-sensitive evaluation process that integrates patient history, physical examination, vital signs, laboratory results, and scoring systems such as the Sequential Organ Failure Assessment (SOFA) and the Systemic Inflammatory Response Syndrome (SIRS) criteria [11,12]. Additional data, including hemodynamic monitoring, imaging studies, and microbiological analyses, are often required to identify the underlying etiology [9,10].

Despite the availability of these diagnostic tools, early recognition of circulatory shock remains challenging in practice. The large volume of heterogeneous clinical data generated in ICU environments, combined with the high workload faced by healthcare professionals, can delay decision-making and reduce diagnostic accuracy [13,14]. In this context, machine learning has become a promising approach to support clinical decision-making by identifying complex patterns in large datasets and enabling earlier prediction of critical conditions [15].

## 2. RELATED WORKS

Several limitations can be identified in the current scientific literature on machine learning applications for circulatory shock. First, there is an imbalance in research focus, with most studies concentrating on septic shock (or sepsis), while other types, particularly hypovolemic shock, remain underrepresented. Second, many models are designed to detect or predict a single type of shock or identify shock without differentiating its underlying etiology, rather than addressing the problem as a multiclass classification task.

A representative example is the study by Jentzer et al. (2025) [16], which developed several machine learning models to predict the onset of shock in ICU patients using routinely collected vital signs from the first four hours after ICU admission. The best-performing model, XGBoost, achieved an AUROC of 0.82 and an accuracy of 81% for predicting shock within the following four hours. However, the models treated shock as a single outcome and did not distinguish between its different etiologies.

Beyond general shock prediction, the literature has predominantly focused on septic shock. As comprehensively reviewed by Zubair et al. (2025) [17], numerous machine learning and deep learning approaches have been proposed for the early detection and prediction of sepsis and septic shock. For example, Agor et al. (2022) [18] developed a temporal pattern mining framework that analyzes electronic health records (EHRs) to identify clinically relevant temporal patterns associated with the onset of septic shock. Longitudinal clinical data were transformed into multivariate state sequences, from which temporal patterns were extracted and used as predictive features for conventional machine learning classifiers, including Support Vector Machines (SVM) and Logistic Regression.

Building on these approaches, Kim et al. (2023) [19] introduced DeepSEPS, a deep learning-based system for the real-time prediction of both sepsis and septic shock using EHR data. Developed from retrospective data collected between 2010 and 2019 and externally validated on patients admitted between 2020 and 2021, the model incorporated demographic variables, vital signs, and laboratory measurements. Compared with widely used clinical scoring systems, including NEWS, SOFA, and qSOFA, DeepSEPS consistently achieved superior performance, reaching an AUROC of 0.9346 at the onset of sepsis and septic shock.

Similarly, Misra et al. (2021) [20] developed a machine learning-based clinical decision support system to predict the progression from sepsis to septic shock within the first six hours of ICU admission. Using EHR data, the authors compared eight machine learning algorithms, including Random Forest, XGBoost, Support Vector Machine, and Logistic Regression.

Random Forest achieved the best performance, with an AUROC of 0.9483, a sensitivity of 83.9%, and a specificity of 88.1%.

Although septic shock has received the greatest attention, research on machine learning applications for cardiogenic shock has also expanded considerably. As summarised by Vallabhajosyula et al. (2026) [21], several machine learning approaches have demonstrated promising predictive performance. For instance, Chang et al. (2022) [22] developed an EHR-based prediction model in which XGBoost achieved an AUC of 0.87 for predicting cardiogenic shock at least two hours before its clinical onset. The model also showed good interpretability, with its most influential predictors aligning with established clinical evidence.

Likewise, Hu et al. (2024) [23] proposed CShock, a deep learning-based risk stratification model for the early prediction of cardiogenic shock in patients admitted to the cardiac intensive care unit. Developed using demographic, clinical, laboratory, and imaging data from EHRs and externally validated on an independent cohort, CShock achieved AUROCs of 0.821 and 0.800 in the development and validation cohorts, respectively, demonstrating good generalizability and potential as an automated early warning system.

In contrast, the early prediction of hypovolemic shock has received considerably less attention. One of the few available studies was conducted by Zhao et al. (2022) [24], who proposed an XGBoost-based model for the early prediction of traumatic hemorrhagic shock, a major subtype of hypovolemic shock, using EHR data. Their stepped feature strategy progressively incorporated vital signs, routine blood tests, and blood gas analysis as data became available. The proposed model achieved an AUROC of up to 0.968 for predicting traumatic hemorrhagic shock one hour before clinical recognition and demonstrated good external generalizability.

While the studies discussed above focus on either general shock detection or individual shock subtypes, only one study has attempted to combine shock detection with etiological differentiation. Nemeth et al. (2021) [25] developed a machine learning-based decision support system for prehospital tactical combat casualty care using routinely collected vital signs. The authors trained logistic regression models on ICU electronic health record data to identify patients at risk of circulatory shock and distinguish among septic, cardiogenic, and hypovolemic shock up to 90 minutes before clinical recognition. At the time of shock recognition, the models achieved sensitivities of 78% for general shock, 77% for septic shock, 75% for cardiogenic shock, and 60% for hypovolemic shock. However, the authors did not formulate the task as a multiclass classification problem nor report multiclass evaluation metrics, making direct comparison with multiclass approaches difficult.

Beyond these methodological limitations, the studies reviewed above rely on markedly heterogeneous sources. Only two employed a publicly available critical care database: Hu et al. (2024) [23] developed CShock using MIMIC-III, with external validation on an independent NYU Langone cohort, while Zhao et al. (2022) [24] combined the PLA General Hospital Emergency Rescue Database (PLAGH-ERD), an institutional trauma registry, with MIMIC-III as an external validation set. Nemeth et al. (2021) [25] likewise trained their shock differentiation algorithm on MIMIC before refining it with Mayo Clinic ICU data. The remaining studies relied exclusively on institutional electronic health record systems, including Jentzer et al. (2025) [16], who drew on Mayo Clinic data; Agor et al. (2022) [18], who relied on the Christiana Care Health System in Wilmington, Delaware; Kim et al. (2023) [19], who sourced data from a South Korean tertiary hospital; Misra et al. (2021) [20], who employed the Geisinger Health System; and Chang et al. (2022) [22], who extracted their cohort from the Banner Health system, a 30-hospital regional network in the United States. This heterogeneity in data provenance limits the reproducibility and external comparability of their findings.

Building upon these observations, the present study proposes a machine learning framework for the early prediction and etiological classification of circulatory shock in ICU patients using MIMIC-IV. Specifically, the proposed framework aims to distinguish among septic, hypovolemic, and cardiogenic shock using routinely collected clinical data from the first six hours following ICU admission. To the best of our knowledge, and based on the literature reviewed, this is the first study to apply a multiclass machine learning approach to the early etiological classification of circulatory shock into these three major categories using data from MIMIC-IV. By combining early prediction with multiclass classification on a publicly available, reproducible database, this study addresses the limitations identified in previous research and paves the way for future clinical decision support systems capable of the early identification and etiological classification of circulatory shock.

## 3. METHODOLOGY

An overview of the complete methodological pipeline is presented in Figure 1. The following sections describe each stage in detail.

### 3.1. Database: MIMIC-IV

The database supporting this paper is MIMIC-IV v3.1, one of the most comprehensive and widely used open-access clinical repositories in biomedical research [26–28]. This resource collects detailed and anonymised information on patients treated at Beth Israel Deaconess Medical Centre in Boston between 2008 and 2019. Overall, it includes data from more than 65,000 patients admitted to the ICU. Once the MIMIC-IV dataset was downloaded, PostgreSQL was installed on a personal computer to manage, visualise, and query the data.

### 3.2. Data extraction

Patients were initially selected based on an ICU length of stay of at least 8 hours, in order to ensure sufficient clinical data availability and enhance comparability among cases. The study population was restricted to adult patients (≥18 years). To ensure accurate classification of patients by type of circulatory shock within the initial population, four cohorts were constructed based on the codes and definitions provided in MIMIC-IV, incorporating both diagnostic classifications (ICD-9/ICD-10) and the extraction of relevant physiological and laboratory variables.

#### 3.2.1. Hypovolemic shock patient cohort

For the hypovolemic shock cohort, specific codes available in MIMIC-IV were used, including those corresponding to hypovolemic shock (R571) and traumatic or hemorrhagic shock (T794), the latter considered a subtype of hypovolemic shock. In addition, patients with a diagnosis of hypovolemia (27652) and a first ICU stay longer than 24 hours were included, even in the absence of an explicit hypovolemic shock code. This criterion was applied to ensure the inclusion of clinically relevant cases with sufficient evolution, capturing severe hypovolemia associated with significant hemodynamic instability that may correspond to unrecorded hypovolemic shock.

Furthermore, all patients meeting criteria for other shock cohorts (septic or cardiogenic) were excluded to avoid overlap and ensure that each patient was assigned exclusively to the appropriate pathophysiological category. Patients who developed complications during hospitalisation, such as sepsis (identified through the corresponding ICD codes), were also excluded to preserve the initial hemodynamic state and avoid confounding clinical conditions that could lead to a different shock profile

#### 3.2.2. Septic shock patient cohort

For the septic shock cohort, only patients with a specific diagnosis code for septic shock in MIMIC-IV were included (78552 and 99802, the latter corresponding to postoperative septic shock). Patients with sepsis without shock, identified using the corresponding ICD codes [29], were excluded, as they represent a less severe clinical condition with a different hemodynamic profile, their inclusion would have introduced undesired heterogeneity into the cohort. As in the previous cohort, all patients meeting inclusion criteria for other shock categories (hypovolemic or cardiogenic) were excluded to avoid overlap and to ensure that each individual was assigned exclusively to the pathophysiological category that best represented their initial clinical state.

#### 3.2.3. Cardiogenic shock patient cohort

For the cardiogenic shock cohort, only the specific code for cardiogenic shock available in MIMIC-IV (78551) was used to identify patients in this category. Individuals who simultaneously met criteria for other shock cohorts (septic or hypovolemic) were excluded to avoid overlap and to ensure that each case represented a clearly defined pathophysiological pattern. Additionally, patients who developed sepsis during hospitalisation were excluded, as this complication may significantly alter the initial hemodynamic profile and lead to a mixed clinical state that does not correspond to primary cardiogenic shock.

#### 3.2.4. Cohort of patients without shock

To reduce the risk of including patients with unrecognised shock, the non-shock cohort was defined based on the absence of vasoactive drug use during the first ICU stay. This included both vasopressors (norepinephrine, epinephrine, phenylephrine, vasopressin, and dopamine) and inotropes (dobutamine and milrinone). The absence of vasoactive therapy serves as a robust proxy for hemodynamic stability, as these agents are routinely administered to manage sustained hypotension, hypoperfusion, or impaired cardiac contractility [30,31]. Additionally, patients with diagnoses of sepsis or other types of shock not included in the predefined groups (cardiogenic, hypovolemic, or septic) were excluded. These cases were identified using ICD codes corresponding to conditions such as anaphylactic, toxic, neurogenic, obstetric, drug-induced, burn-related, or unspecified shock.

### 3.3. Preprocessing and statistical analysis of clinical data

To standardise the analysis, only the first ICU admission for each patient was considered for the extraction and evaluation of clinical variables. All variables were analysed within the first 6 hours following ICU admission, with the aim of capturing the early phase of shock. For each patient, the mean value of all observations recorded during this time window was computed, with the exception of urine output.

A diverse set of variables was analysed and grouped into four main categories: physiological variables, including mean arterial pressure (MAP), diastolic and systolic blood pressure, heart rate, respiratory rate, oxygen saturation ($SpO_2$), urine output, and temperature; derived clinical indices, including the modified shock index (MSI) [32] defined as the ratio of heart rate to mean arterial pressure (HR/MAP) and the SaFi ratio [33], defined as $SpO_2/FiO_2$ ($FiO_2$: fraction of inspired oxygen); laboratory variables, including lactate, bicarbonate, blood pH, white blood cell count (WBC count), hematocrit, hemoglobin, creatinine, anion gap, and blood urea nitrogen (BUN); and demographic variables, including age and gender.

Before analysis, all variables underwent preprocessing to remove inconsistent values, extreme outliers, and physiologically implausible records in order to improve data quality and reliability. The plausible value ranges and corresponding item IDs used for this filtering process are detailed in Table S1 of the Supplementary Material. Initially, a descriptive analysis was performed for all variables included in the study. Mean, median, quartiles, minimum values, and maximum values were calculated without considering missing values (NA). A longitudinal analysis of the main clinical variables was then conducted to characterise the temporal evolution of hemodynamic status during the first five days of the initial ICU stay. The variables examined included mean arterial pressure, heart rate, respiratory rate, and modified shock index. Mean values were calculated in two-hour intervals for each patient and subsequently averaged by cohort, and visualisations included 95% confidence intervals.

The normality of continuous variables was evaluated using the Kolmogorov–Smirnov test and Q-Q plots. Since the variables did not follow a normal distribution, group comparisons were performed using non-parametric methods. A Kruskal-Wallis test was applied as a global analysis, followed by Dunn's post-hoc test with Holm correction. Effect sizes were quantified using epsilon-squared ($\varepsilon^2$) for the global comparisons and rank-biserial correlation (r) for pairwise comparisons. For the categorical variable gender, global comparisons were performed using the $\chi^2$ test, with effect size quantified using Cramér's V. Pairwise comparisons were performed using Fisher's exact test when appropriate, with effect sizes quantified using the Phi coefficient ($\varphi$). Post-hoc comparison results were represented using heatmaps showing statistical significance and effect size magnitude (Figure S1 and Figure S2 of the Supplementary Material).

### 3.4. Data imputation

A missing-at-random (MAR) mechanism [34] was assumed for the missing data, as the clinical context and the nature of the data indicate that the availability of certain variables depends on observed patient characteristics, such as disease severity, ICU stay progression, or clinical decision-making.

Missing values were quantified for each variable by calculating the percentage of absent data. Due to the high degree of incompleteness observed, several imputation strategies were considered and compared, ranging from lower to higher methodological complexity. First, a simple median-based imputation was applied within each group. Subsequently, more advanced imputation methods were evaluated, including the k-nearest neighbours (k-NN) algorithm with k values of 1, 3, 5, 7, and 9 [35]. Finally, missing data were imputed using the Multiple Imputation by Chained Equations (MICE) approach [36], implemented as a single iterative imputation procedure.

To compare the different imputation strategies, four multiclass classification models were trained using default hyperparameters: Random Forest, Logistic Regression, XGBoost, and a multilayer perceptron with a 64:32 architecture. This comparison was performed as a preliminary step, using a separate balanced subset of 2,400 patients (600 individuals from each clinical cohort: septic shock, hypovolemic shock, cardiogenic shock, and non-shock controls), independent of the dataset used for final model development and evaluation. Each model was evaluated using all considered imputation methods (median, k-nearest neighbours, and MICE), while median-based imputation was used as the baseline, as model training was not feasible without handling missing values. Model performance was assessed using stratified 10-fold cross-validation, with each observation serving once as validation data and nine times as training data across the 10 folds. Performance metrics were averaged across the 10 folds to provide robust performance estimates without requiring an independent test set.

### 3.5. Model training and validation

To develop the models, the data were loaded into Jupyter Notebook (7.4.5) using Python 3.12. The NumPy (2.3.3) and Pandas (2.3.2) libraries were used for data processing, Scikit-learn (1.7.1) for model development and evaluation, and Matplotlib (3.10.6) for data visualisation and result presentation. For this stage, the full clinical dataset extracted from MIMIC-IV was used, independent of the 2,400-patient subset used for the imputation comparison described above. A two-stage cascade model was defined for each evaluated algorithm. In the first stage, a binary classifier (Model 1) was trained to distinguish between shock and non-shock patients. If Model 1 predicted the absence of shock, the patient was classified directly as non-shock, if shock was detected, the case was passed to the second stage, where the same algorithm was retrained as a multiclass classifier (Model 2) to differentiate among the three shock subtypes (septic, hypovolemic, and cardiogenic). Thus, four independent cascade architectures were evaluated, one for each machine learning algorithm (Random Forest, Logistic Regression,

MLP 128:64, and XGBoost). For the MLP, a 128:64 architecture was used, following hyperparameter optimisation distinct from the 64:32 configuration applied in the imputation comparison.

The dataset was split into training (75%) and test (25%) subsets. Within the training set, the MICE imputer was fitted and used to transform the training data; SMOTE (Synthetic Minority Over-sampling Technique) was then applied to the minority classes [37], while the majority class (non-shock) was balanced through random undersampling [38], resulting in a final training size of 3,000 patients for each shock subtype. For the test set, the previously trained MICE imputer was applied in transform-only mode to prevent data leakage. The imputed test set then underwent a guided subsampling procedure designed to approximate literature-reported shock prevalence [10], constrained by the limited availability of minority classes (cardiogenic and hypovolemic shock) within MIMIC-IV. Target proportions were based on literature-reported prevalence: 67% for non-shock, 18.2% for septic shock, 5.2% for hypovolemic shock, and 4.9% for cardiogenic shock. This yielded a curated test cohort of 2,300 patients.

After defining the training and test sets, hyperparameter optimisation was performed separately for each cascade model using GridSearchCV combined with stratified cross-validation on the training data only. In the first stage, 10-fold stratified cross-validation was applied, with hyperparameters selected based on the highest AUROC achieved by the binary classifier (Model 1). In the second stage, 10-fold stratified cross-validation was also used, with hyperparameters selected based on the highest macro F1-score achieved by the multiclass classifier (Model 2). Final cascade models were then trained using the optimised hyperparameters. Performance was subsequently evaluated on the independent test cohort across four domains: discrimination, assessed via the AUROC of the first-stage binary classifier [39]; classification performance, assessed via F1-score, precision (PPV), sensitivity (TPR), and negative predictive value (NPV); specificity, reported as both macro and per-class specificity; and the confusion matrix summarising per-class prediction patterns. Each classification metric was calculated for every class and macro-averaged to provide a comprehensive assessment of overall cascade performance. Additional interpretability analyses were conducted using SHAP (SHapley Additive exPlanations) 0.48.0 to identify the most influential predictors for each classification task [40,41].

Given the importance of interpretability in clinical settings, SHAP values were calculated for each of the four optimised classifiers (Random Forest, XGBoost, Logistic Regression, and MLP 128:64) to improve model transparency and facilitate clinical understanding of the predictions. For each model, the ten most relevant variables were visualised across the four classification tasks considered: the binary classification of shock versus non-shock, and the three one-vs-rest multiclass tasks corresponding to hypovolemic shock, cardiogenic shock, and septic shock.

## 4. RESULTS

### 4.1 Statistical analysis

After defining the cohorts, a study population of 878 patients with hypovolemic shock, 603 patients with cardiogenic shock, 1,916 patients with septic shock, and 29,510 patients without shock was obtained. The demographic characteristics of the cohorts were generally comparable. Age distribution was similar across groups, with overlapping means and standard deviations, suggesting the absence of clinically meaningful differences in age between patients with and without shock. In contrast, some differences were observed in gender distribution. The most notable variation was found in the cardiogenic shock group, which included a higher proportion of male patients, possibly due to the later onset of cardiovascular disease in women [42].

The baseline clinical characteristics of the study cohorts, summarised in Table 1, showed that the non-shock group maintained adequate hemodynamic stability, presenting the lowest MSI values. Cardiogenic shock was characterised by marked pump dysfunction, with hypotension, elevated lactate levels, metabolic acidosis, and signs of renal impairment, accompanied by an increased MSI. Hypovolemic shock presented with tachycardia, hypotension, moderate reduction in urine output, and hematological alterations compatible with volume loss or hemodilution, also showing elevated MSI values. Finally, septic shock represented the most severe condition, characterised by the most pronounced combination of hypotension, tachycardia, respiratory deterioration, marked reduction in urine output, and metabolic acidosis. These findings were reflected in the highest MSI values among all groups, indicating severe hemodynamic decompensation.

These clinical differences were further supported by the statistical analysis presented in Table 2. All clinical and demographic variables demonstrated significant global differences among the four groups ($p < 0.001$ for all comparisons), while pairwise comparisons are provided in Figure S1 and Figure S2 of the Supplementary Material. Despite this consistent statistical significance, the magnitude of the observed differences was generally modest. Systolic blood pressure and the modified shock index displayed the highest discriminative capacity, with moderate effect sizes ($\varepsilon^2$: 0.0713 and 0.0635, respectively). All remaining variables exhibited small or very small effect sizes, as detailed in Table 2. Collectively, these findings indicate that although the differences between groups are statistically robust, likely owing in part to the large sample size, their clinical relevance should be interpreted cautiously because the observed effect sizes are relatively limited.

Regarding temporal evolution, longitudinal analyses revealed clear physiological differences between shock types during the first days of ICU stay (Figure 2). Septic shock showed the greatest physiological disturbance, with persistently elevated heart rate, respiratory rate, and MSI values, although a gradual stabilisation trend was observed over time, suggesting a response to treatment and supportive interventions. Hypovolemic and cardiogenic shock presented intermediate patterns, with moderate hemodynamic instability and lower variability compared with septic shock, indicating a less severe and more homogeneous clinical evolution during the first ICU days. In contrast, the non-shock group maintained the greatest physiological stability throughout the observation period. These longitudinal analyses also demonstrated that the most relevant physiological changes occurred during the first hours after ICU admission, before patients reached a state of relative stabilisation. This pattern was especially evident in shock groups, where early alterations in heart rate, mean arterial pressure, respiratory rate, and MSI reflected the phase of greatest hemodynamic instability.

## 4.2 Data imputation

As shown in Table 3, the proportion of missing values varied across variables and clinical groups. The highest levels of missingness were observed for urine output (ml/kg/h), the SaFi index, and several laboratory variables. Conversely, basic physiological variables exhibited low rates of missing data across all groups, generally below 3-4%. Age and gender were complete variables, with no missing values detected in any of the analysed groups.

In contrast, variables such as urine output (ml/kg/h) presented a high percentage of missing data, especially in the group without shock (43.19%), although it was also notable in the different types of shock. Similarly, the SaFi index showed the highest proportion of missing values in all groups, reaching 70.42% in patients without shock and remaining above 40% in the different types of shock. Regarding laboratory variables, high percentages of missing values were observed for blood lactate and pH, particularly in the group without shock, while in the shock groups, the incompleteness was lower, although still considerable. Other analytical variables showed moderate percentages of missing values, generally ranging between 8% and 28%, with greater data availability in patients with shock than in the group without shock.

Once missing values had been identified, they were imputed using the methods previously described. k-NN imputation was implemented using a specific value of k = 7, selected after evaluating different k values through cross-validation, as it demonstrated the best overall performance. After training the 12 multimodal models (four for each imputation strategy) and completing the cross-validation process, the comparison of the results based on macro F1-score, macro precision, macro sensitivity, and macro specificity did not reveal any imputation strategy that was clearly superior in terms of overall performance. As shown in Table 4, the values obtained were highly similar across the different imputation strategies considered.

Median imputation and MICE yielded marginally better results for some models and metrics. Specifically, the best performance with median imputation was achieved by RF and XGB (macro F1-score: 0.49 ± 0.02), whereas RF achieved the highest macro F1-score with MICE (0.50 ± 0.03). However, these differences were small and unlikely to be clinically relevant. Therefore, MICE was selected as the imputation strategy for subsequent analyses, owing to its widespread use in clinical datasets under the missing at random (MAR) assumption and its ability to preserve multivariate relationships among the analysed variables [43,44].

## 4.3 Model training and validation

After splitting the dataset, the training set consisted of 75% of the patients, while the remaining 25% was reserved for independent model evaluation. As a result of applying the balancing procedure to the training set, a balanced class distribution was obtained, with a total of 3,000 patients per shock type used for model training.

The test set, comprising 2,300 patients, was obtained through a guided subsampling procedure designed to approximate the shock prevalence reported in the literature. The final class proportions in the test set were 67% (1,541 patients) for the no-shock group, matching the literature-based target; 21.1% (485 patients) for septic shock, representing a 2.9-percentage-point increase over the 18.2% target; 6.1% (141 patients) for hypovolemic shock, a 0.9-percentage-point increase over the 5.2% target; and 5.8% (133 patients) for cardiogenic shock, also a 0.9-percentage-point increase over the 4.9% target. Overall, the resulting distributions were highly similar to the target proportions. The slight deviations observed are explained by the fact that this paper did not include other less frequent types of circulatory shock beyond the three major categories considered.

After defining the final training and test sets, Grid Search combined with stratified cross-validation was applied to optimise the hyperparameters of the evaluated cascade models. However, hyperparameter tuning resulted in only limited performance improvements. Random Forest and Logistic Regression showed negligible changes compared with their baseline configurations, whereas the MLP exhibited the most noticeable benefit from optimisation. The best-performing MLP configuration increased the size of the hidden layers from 64:32 to 128:64 neurons. XGBoost achieved only marginal improvements, indicating that the models were already operating close to their optimal performance before tuning.

Overall, Random Forest and XGBoost achieved highly comparable performance across all evaluated metrics (Table 5). Random Forest was selected for further interpretation because it showed a slightly better ability to identify hypovolemic shock, the most complex classification task in the study, while maintaining performance comparable to XGBoost across the remaining categories. The analysis revealed notable differences in performance according to shock etiology. The best results were obtained for the no shock category, with F1-scores of 0.80-0.82 across all models, indicating that the algorithms were highly effective at distinguishing patients who did not meet shock criteria from those who did. This finding is consistent with the strong performance observed in the first stage of the cascade system and supports the robustness of the proposed approach for early shock detection.

Among the shock subtypes, septic shock was identified most accurately, achieving F1-scores between 0.52 and 0.67 and sensitivities up to 0.76. Cardiogenic shock showed intermediate performance, with F1-scores ranging from 0.29 to 0.59, indicating that some true cases were still misclassified into other categories despite a low rate of false positives. In contrast, hypovolemic shock was the most difficult class to identify, with sensitivities of only 0.21-0.34 and F1-scores of 0.17-0.33 across models.

As shown in Table 6, across all algorithms, the shock vs. no-shock classification was driven by a consistent and clinically coherent set of variables. Systolic blood pressure, mean arterial pressure, and the modified shock index dominated the rankings in every model, reflecting their direct role in the hemodynamic definition of shock. These were accompanied by metabolic markers of end-organ hypoperfusion, namely blood urea nitrogen, creatinine, and bicarbonate, as well as variables capturing the broader physiological response to circulatory failure, including respiratory rate, heart rate, and WBC count. The high degree of agreement between algorithms in this task indicates that the hemodynamic and metabolic signature of shock is sufficiently robust to be captured consistently regardless of the modelling approach.

In contrast to the binary classification, the etiology tasks showed greater divergence in feature ordering across models, while still converging on the same broad physiological categories (Table 6). For the septic shock vs. other shocks task, all four models assigned high importance to variables associated with systemic inflammatory response, particularly WBC count, temperature, and bicarbonate, combined with global hemodynamic parameters such as MAP and the modified shock index, and oxygen-carrying variables including hemoglobin and hematocrit, reflecting the dual hemodynamic and inflammatory phenotype that characterises septic shock. For the hypovolemic shock vs. other shocks classification, the models placed greater relative importance on renal and hematological markers such as creatinine, hematocrit, and hemoglobin. Finally, in the cardiogenic shock vs. other shocks classification, hematological variables and metabolic markers such as anion gap and bicarbonate were consistently highlighted together with systolic blood pressure and MAP.

Across the four classification tasks, Random Forest, Logistic Regression, MLP, and XGBoost identified largely overlapping sets of influential predictors. The most important variables consistently belonged to the same physiological domains, including hemodynamic status, metabolic disturbances, renal function, inflammatory response, and respiratory compensation. Although the relative ranking of individual variables varied among algorithms, a substantial degree of agreement was observed regarding the variables contributing most strongly to model predictions.

#### 4.3.1 Detailed Evaluation of the Random Forest Model

Random Forest was selected as the representative model for in-depth evaluation because its performance was comparable to that of the best-performing models, while showing a slight advantage in classifying hypovolemic shock. Table 7 presents the row-normalised confusion matrix for the Random Forest cascade, highlighting the model's class-specific classification performance. This matrix revealed distinct patterns of misclassification. All three types of circulatory shock were frequently misclassified as no shock: 28.6% of cardiogenic shock cases, 46.1% of hypovolemic shock cases, and 17.5% of septic shock cases. Septic shock was sometimes misclassified as cardiogenic or hypovolemic shock. In contrast, among no-shock cases, misclassifications were predominantly toward septic shock (14.5%), while confusion with cardiogenic and hypovolemic shock occurred less frequently.

To complement this analysis, SHAP was used to interpret the model's predictions. The corresponding beeswarm plots (Figure 3) display the contribution of each feature to the model predictions across all patients, with colour indicating the feature value (red: high, blue: low). Although the exact ordering of features varied between models, similar patterns were observed across algorithms. In all cases, the most influential variables belonged to the same clinically relevant physiological domains, including hemodynamic status, renal function, metabolic disturbances, inflammatory response, and respiratory compensation.

For the shock detection task, the beeswarm plot revealed that hemodynamic instability markers dominated model predictions, with systolic blood pressure and MSI standing out as the most influential features. Other highly influential variables included BUN, creatinine, bicarbonate, and pH, indicating that renal function and metabolic status also contributed substantially to model predictions. Lower values of systolic blood pressure and mean arterial pressure (MAP), together with higher values of MSI, creatinine, and BUN, were associated with an increased probability of shock.

For the septic shock classification task, the most influential features were MSI, temperature, respiratory rate, and WBC count. Higher values of MSI, temperature, respiratory rate, and WBC count contributed positively to septic shock predictions, whereas lower hematocrit and hemoglobin values were associated with an increased probability of septic shock relative to the other shock subtypes. In the cardiogenic and hypovolemic shock models, the most influential variables included temperature, hemoglobin, hematocrit, anion gap, creatinine, BUN, MSI, and respiratory rate. Lower values of temperature, hemoglobin, and hematocrit, together with higher anion gap values, were associated with an increased probability of cardiogenic shock relative to the remaining shock categories. In contrast, lower values of creatinine, BUN, anion gap, MSI, and respiratory rate were associated with an increased probability of hypovolemic shock relative to the other shock subtypes.

## 5. DISCUSSION

Given the results presented above, machine learning algorithms were able to differentiate patients with shock from those without shock with reasonable accuracy. However, distinguishing between different shock etiologies proved considerably more difficult, particularly for hypovolemic and cardiogenic shock, which were frequently misclassified. In contrast, septic shock was identified with comparatively better performance across all evaluated models.

One possible explanation is that septic shock exhibits a stronger and more distinctive biological signature, characterised by marked inflammatory, hemodynamic, and metabolic alterations. Conversely, hypovolemic shock may present a weaker or less specific physiological profile, making it more difficult for machine learning algorithms to distinguish it from other shock states. In addition, the difficulty in identifying hypovolemic shock may be related to its often favourable response to fluid resuscitation [30], resulting in physiological parameters that more closely resemble those of non-shock patients by the time data are collected. Consistent with this explanation, Table 7 shows that nearly half of hypovolemic shock cases were misclassified as non-shock, compared with fewer than one-third of cardiogenic shock cases and only a small minority of septic shock cases. Furthermore, the lower performance observed for hypovolemic shock may also be influenced by the criteria used to define this cohort.

These findings are consistent with the limited evidence currently available. In particular, Nemeth et al. reported lower predictive performance for cardiogenic and especially, hypovolemic shock than for septic shock when developing separate binary machine learning models for each shock subtype [25]. This recurring pattern may indicate that the lower classification performance observed for these subtypes is attributable, at least in part, to their intrinsic clinical overlap and heterogeneity rather than solely to limitations of the machine learning models themselves.

Although direct comparisons remain difficult due to the limited number of multiclass studies available, most machine learning studies in this field have instead focused on binary tasks. Within this literature, septic shock generally appears to achieve the most favourable predictive performance [45], likely reflecting its more distinctive inflammatory and physiological profile. In contrast, results for cardiogenic shock are more variable, with some studies reporting good discrimination while others achieve only moderate performance [46]. The evidence regarding hypovolemic shock remains limited, which may itself reflect the greater difficulty of identifying this subtype compared with other forms of shock.

To better understand the factors underlying these performance differences, the feature importance and SHAP analyses of the Random Forest model, which achieved the best overall performance, were examined. Most of the predictors used for shock detection and shock etiology classification were consistent with current scientific knowledge. For shock identification, the most influential variables included systolic blood pressure, MAP, MSI, BUN, and creatinine. Lower systolic blood pressure and MAP values, together with higher MSI, BUN, and creatinine levels, were associated with an increased probability of shock. These findings are physiologically coherent, as shock is characterised by circulatory failure and inadequate tissue perfusion, leading to hemodynamic instability and organ dysfunction. The prominence of renal function markers such as BUN and creatinine is also expected, given the sensitivity of the kidneys to reduced perfusion during states of circulatory compromise [9,10].

When examining shock etiology classification, the identified predictors also showed substantial agreement with the physiological mechanisms described in the literature. In the septic shock versus other shocks task, high MSI, temperature, WBC count, and respiratory rate were important predictors. These variables are consistent with the systemic inflammatory response that characterises sepsis and correspond closely to the traditional SIRS criteria [47], while the elevated MSI reflects the associated hemodynamic instability observed in septic shock.

However, the importance assigned to lower hemoglobin and hematocrit values is less straightforward to explain from a physiological perspective. Current evidence does not indicate that septic shock patients should inherently present lower hemoglobin or hematocrit levels than patients with other forms of shock. A more plausible explanation is that these variables may be capturing treatment-related effects. Because patients with septic shock frequently receive aggressive fluid resuscitation during the early stages of management [9,47,48], lower hemoglobin and hematocrit values may reflect hemodilution secondary to intravenous fluid administration rather than a direct biological characteristic of septic shock itself.

Regarding the classification of cardiogenic shock versus other shock etiologies, one of the most notable findings was the high importance assigned to hematocrit. From a physiological perspective, there is no established mechanism by which cardiogenic shock would consistently present with higher hematocrit values than other major shock etiologies. A more plausible explanation is that the model may be capturing treatment-related differences rather than intrinsic disease characteristics. Unlike septic or hypovolemic shock, where fluid resuscitation is a central component of management, patients with cardiogenic shock typically receive more restrictive fluid administration [49,50]. Consequently, hemoglobin and hematocrit levels may remain relatively higher, leading the model to associate these variables with cardiogenic shock despite them reflecting therapeutic management rather than the underlying pathophysiology.

The remaining predictors are generally consistent with current knowledge of cardiogenic shock. Higher anion gap values may reflect impaired tissue perfusion and metabolic disturbances, while lower temperature values are consistent with the absence of the pronounced inflammatory response commonly observed in septic shock. Taken together, several of the identified variables, including lower systolic blood pressure, altered MSI values, elevated anion gap, and lower SaFi, are compatible with the classic "wet and cold" profile frequently associated with cardiogenic shock, characterised by both systemic hypoperfusion and pulmonary congestion [49,51]. Nevertheless, these findings should be interpreted cautiously. Although the identified predictors show a degree of physiological coherence, the classification performance for cardiogenic shock was only moderate. Therefore, the importance assigned to individual variables may not necessarily reflect robust pathophysiological markers and should be considered exploratory until validated in external cohorts.

For the classification of hypovolemic shock versus the remaining shock etiologies, the relatively low classification performance achieved for this group should be emphasised. Therefore, the interpretation of feature importance results requires particular caution. For many of the most influential variables, including MSI, respiratory rate, anion gap, and systolic blood pressure, values generally associated with a less severe physiological profile were linked to a higher probability of hypovolemic shock. This finding may be related to the fact that hypovolemic shock appears to exhibit the weakest and least distinctive biological signal among the evaluated shock subtypes. In addition, hypovolemic shock often responds rapidly to fluid resuscitation [52,53], which may attenuate the physiological abnormalities present at the time of data collection and make these patients appear closer to non-shock states.

Another noteworthy finding was the importance assigned to lower hematocrit values, which ranked among the most relevant predictors for hypovolemic shock. While reduced hematocrit may be consistent with blood loss in patients with hemorrhagic hypovolemia, it may also reflect hemodilution resulting from fluid resuscitation. Therefore, this predictor could be influenced by both the underlying condition and treatment-related effects [52,53]. Age also appeared among the most important predictors, with younger age being associated with a higher probability of hypovolemic shock. Although age is not a direct marker of shock physiology, this finding may reflect differences in the demographic profiles of the shock cohorts. In particular, age may help distinguish hypovolemic shock from cardiogenic shock, as patients with cardiogenic shock were generally older in our cohort. This pattern was consistently observed in both the descriptive analysis and the pairwise statistical comparisons between groups.

Taken together, these findings indicate that the models primarily relied on physiologically meaningful variables related to hemodynamic instability, organ hypoperfusion, metabolic disturbances, and systemic inflammation. However, the recurrent importance of hemoglobin and hematocrit across the etiology classification tasks suggests that some predictors may also be capturing treatment-related effects, particularly differences in fluid resuscitation strategies among shock subtypes. This observation highlights the complexity of distinguishing between biological signals arising from the disease process itself and those resulting from clinical management. Furthermore, the interpretation of these findings should also consider the presence of missing data in several variables. Although MICE imputation was used to mitigate this limitation, some influence on model performance cannot be excluded. This issue may be particularly relevant in less severe patients, especially those without shock, who are often monitored less intensively and therefore may present a greater proportion of missing measurements.

Although differentiating shock etiologies remains challenging and some predictors may reflect treatment-related effects rather than underlying pathophysiology alone, and missing data may have influenced model performance despite the use of MICE imputation, the ability to identify shock itself may be of greater clinical relevance. The greatest potential benefit of a machine learning-assisted system may therefore lie in facilitating earlier recognition of circulatory failure rather than in achieving perfect etiological classification [54]. This aspect cannot be evaluated in the present retrospective study, as the temporal relationship between model predictions and clinical diagnosis was not assessed. Nevertheless, the performance achieved for shock detection indicates that these models could provide useful support for clinical decision-making, particularly in the early identification of patients with circulatory failure. By contrast, their ability to reliably differentiate between shock etiologies remains limited and requires further improvement before routine clinical implementation can be considered.

## 6. LIMITATIONS

This study has several limitations that should be considered when interpreting the findings.

First, several design decisions may have affected model performance, related to cohort selection, inclusion criteria, and subtype definitions.

Patients were initially selected based on an ICU length of stay of at least 8 hours. Although this criterion ensured sufficient physiological data for feature extraction, it may have introduced selection bias by systematically excluding patients who died or were transferred before reaching 8 hours of ICU stay. Consequently, the study population may underrepresent the most critically ill patients, potentially limiting the generalizability of the findings.

A second decision, aimed at increasing cohort size, was to broaden the inclusion criteria to encompass patients with hypovolemia who lacked a formal shock diagnosis code, provided they had at least one day of ICU admission. This likely introduced cases with milder disease or limited hemodynamic compromise, potentially diluting the physiological signal of true hypovolemic shock and contributing to the comparatively lower classification performance observed for this subtype.

A third decision relates to the grouping of hemorrhagic and non-hemorrhagic hypovolemia into a single category, despite their potentially distinct underlying mechanisms and physiological profiles. This may have increased heterogeneity within the hypovolemic cohort, making it more difficult for the models to identify a consistent pattern associated with this shock subtype.

Relatedly, the model relied on mean values of vital signs and laboratory variables from the first six hours of ICU admission rather than dynamic time-series data. While this window captures many early physiological alterations relevant to shock recognition, averaging inputs limits applicability for real-time support and may obscure transient but clinically informative patterns. Future work should explore temporal architectures, such as sequence models or sliding-window approaches, capable of operating on streaming data and less dependent on laboratory values that are not always immediately available.

Second, two constraints inherent to the MIMIC-IV database warrant attention. The database precludes accurate determination of shock onset timing, making it impossible to reliably distinguish patients admitted with already established shock from those who developed it during their ICU stay, which limits assessment of true early-detection performance. Moreover, although treatment-related variables and life-support interventions were intentionally excluded, inputs drawn from the first six hours may still partially reflect treatment effects rather than underlying pathophysiology alone, particularly in hypovolemic shock, where prompt fluid resuscitation can rapidly normalise many physiological abnormalities. Cohort assignment also relied on retrospective definitions derived from diagnostic codes and clinical criteria available in MIMIC-IV; as no universally accepted gold standard exists for retrospective shock classification, some degree of misclassification cannot be excluded.

Third, the high prevalence of missing data represents a potential source of bias. Although MICE is a well-validated and widely used approach, any imputation method carries the risk of introducing artificial correlations or altering variable distributions, with downstream effects on model training, feature importance estimates, and interpretability.

Fourth, the absence of external validation on independent datasets or across different institutions means that generalizability remains uncertain. Future studies should prioritize external validation in diverse clinical settings, as well as prospective evaluation, to establish the model's robustness and potential clinical utility.

## 7. CONCLUSIONS

This study demonstrates the feasibility of applying machine learning models to the early detection and classification of circulatory shock in critically ill patients. The longitudinal analysis of key clinical variables highlighted substantial physiological alterations during the first hours following ICU admission, before patients reached a state of relative stabilization, underscoring the importance of intensive monitoring during this critical period. Among the evaluated algorithms, the Random Forest model achieved the most balanced overall performance, showing a reasonable capacity to identify the presence of shock. Classification was most accurate for the no-shock class and, to a lesser extent, for septic shock, whereas hypovolemic shock proved the most difficult subtype to classify correctly, likely due to its less distinctive physiological profile and its substantial clinical overlap with other shock etiologies.

The variables with the greatest predictive importance were largely consistent with those reported in the clinical literature, supporting the physiological plausibility and interpretability of the models. Hemodynamic variables, markers of organ hypoperfusion, metabolic disturbances, and indicators of systemic inflammation emerged as the most influential predictors across tasks. Nevertheless, the recurrent importance of hemoglobin and hematocrit in the etiology classification models shows that some predictors may also reflect treatment-related effects, particularly differences in fluid resuscitation strategies among shock subtypes, rather than underlying pathophysiology alone. This observation highlights the difficulty of disentangling disease-specific biological signals from the physiological consequences of clinical management in retrospective datasets.

Overall, the results support the potential use of machine learning as a clinical decision-support tool in the ICU, particularly for the identification of circulatory shock. However, further improvements are required before reliable differentiation between shock etiologies can be achieved. Future research should focus on improving discrimination between shock subtypes, developing real-time monitoring systems capable of capturing the temporal evolution of the patient's condition, and validating the proposed approach in independent datasets and diverse clinical settings. The incorporation of treatment-related variables and organ support interventions may further improve model performance and help distinguish underlying pathophysiological mechanisms from the effects of clinical management.

## Declaration of competing interest

The authors declare that they have no known competing financial interests or personal relationships that could have appeared to influence the work reported in this paper.

## CRediT authorship contribution statement

Jaume Aguiló Piña: Conceptualisation, Methodology, Software, Data curation, Formal analysis, Investigation, Visualisation, Writing - original draft.

Laia Subirats: Conceptualisation, Methodology, Writing - review & editing, Supervision, Project administration.

Aina Frau-Pascual: Conceptualisation, Methodology, Writing - review & editing, Supervision, Project administration.

Jordi Morillas Perez: Conceptualisation, Methodology, Validation, Writing - review & editing

Alba Gorriz: Software, Data Curation, Writing - review & editing

Rudys Magrans Nicieza: Methodology, Writing - review & editing

## Acknowledgments

This research was funded by the "Ayudas para contratos para la formación de doctores y doctoras en empresas y otras entidades (Doctorados Industriales) 2025 (DIN2025-014707)".

## Data availability

The data used in this study are derived from the MIMIC-IV database, which is publicly available on PhysioNet (https://physionet.org/content/mimiciv/). Access requires completion of the CITI "Data or Specimens Only Research" training course and execution of a data use agreement, in accordance with PhysioNet's credentialed access policy. The code developed for data extraction, preprocessing, and model training is available from the corresponding author upon reasonable request.

**Table 1.** Baseline characteristics and physiological/laboratory variables by shock subtype. Continuous variables represent the mean of all measurements recorded during the first 6 hours of ICU admission and are presented as median [interquartile range]. Categorical variables are presented as n [%]. BP: blood pressure; MAP: mean arterial pressure; BUN: blood urea nitrogen; WBC: white blood cells.

| *Variable* | ***No shock (N** = 29510**)*** | ***Cardiogenic shock (N**= 603)* | ***Hypovolemic shock (N**= 878)* | ***Septic shock (N**= 1916**)*** |
|---|---|---|---|---|
| Age | 64 [51-76] | 70 [60-80] | 65 [52-79] | 67 [56-80] |
| Male Gender | 15811 [53.0 %] | 352 [59.4 %] | 480 [53.2 %] | 1042 [53.9 %] |
| Temperature | 36.78 [36.50-37.03] | 36.51 [36.00-36.89] | 36.69 [36.33-37.00] | 36.80 [36.42-37.23] |
| Heart rate | 82.00 [71.40-94.88] | 85.73 [74.13-99.00] | 87.40 [75.25-100.13] | 93.50 [79.97-108.41] |
| Diastolic BP | 82.43 [74.12-91.43] | 74.00 [67.35-81.62] | 76.12 [68.43-85.00] | 70.17 [64.06-77.47] |
| Systolic BP | 124.00 [112.10-137.09] | 106.21 [97.43-116.23] | 113.13 [103.43-127.54] | 106.14 [98.20-115.20] |
| MAP | 67.22 [59.17-76.17] | 61.13 [53.64-69.50] | 62.50 [54.75-71.20] | 58.43 [51.83-65.67] |
| Oxygen saturation | 97.28 [95.67-98.86] | 97.25 [95.00-99.00] | 97.88 [96.00-99.37] | 96.86 [95.00-98.43] |
| Respiratory rate | 18.00 [15.86-20.70] | 19.75 [16.98-22.85] | 18.57 [16.19-21.14] | 21.00 [17.85-24.60] |
| Urine output | 0.52 [0.28-0.92] | 0.37 [0.15-0.78] | 0.44 [0.22-0.85] | 0.31 [0.14-0.62] |
| SaFi ratio | 196.86 [160.00-237.50] | 152.86 [113.74-193.36] | 186.33 [142.86-208.20] | 160.83 [119.72-198.00] |
| Modified shock index | 1.00 [0.85-1.17] | 1.15 [0.97-1.38] | 1.14 [0.98-1.32] | 1.33 [1.11-1.56] |
| Lactate | 1.65 [1.20-2.30] | 2.50 [1.70-4.20] | 2.00 [1.39-2.96] | 2.20 [1.50-3.60] |
| Hematocrit | 34.30 [29.73-38.55] | 35.95 [30.90-40.78] | 31.70 [27.46-36.06] | 32.00 [27.79-36.20] |
| Bicarbonate | 24.00 [21.13-26.00] | 22.00 [19.00-25.00] | 22.71 [19.50-25.31] | 20.97 [17.50-24.00] |
| pH | 7.38 [7.34-7.43] | 7.33 [7.25-7.38] | 7.35 [7.28-7.40] | 7.33 [7.25-7.39] |
| WBC count | 9.80 [7.35-13.00] | 12.47 [9.00-17.06] | 10.50 [7.80-14.43] | 13.00 [7.80-19.19] |
| Hemoglobin | 14.00 [12.00-16.50] | 17.00 [14.00-20.00] | 14.00 [12.00-16.33] | 15.50 [13.00-18.50] |
| BUN | 11.40 [9.80-12.90] | 11.80 [10.19-13.40] | 10.70 [9.10-12.17] | 10.50 [9.10-11.90] |
| Creatinine | 0.90 [0.70-1.20] | 1.40 [1.00-2.10] | 1.00 [0.70-1.45] | 1.40 [0.90-2.33] |
| Anion gap | 17.00 [12.00-25.00] | 28.00 [19.00-45.42] | 20.00 [13.00-34.00] | 29.00 [18.00-49.00] |

**Table 2.** Statistical comparison of physiological, laboratory, and demographic variables across shock subtypes (Kruskal-Wallis test for continuous variables, chi-square test for sex). Effect sizes are reported as epsilon-squared ($\varepsilon^2$) for continuous variables and Cramer's V for sex, with magnitude categorised according to conventional thresholds (very small, small, moderate and large). WBC: white blood cells.

| Variable | p-value | $\varepsilon^2$/Cramer's V - Effect size magnitude |
|---|---|---|
| Temperature | < 0.001 | 0.0052 (Very small) |
| Heart rate | < 0.001 | 0.0182 (Small) |
| Systolic blood pressure | < 0.001 | 0.0713 (Moderate) |
| Diastolic blood pressure | < 0.001 | 0.0342 (Small) |
| Mean arterial pressure | < 0.001 | 0.0566 (Small) |
| Oxygen saturation ($SpO_2$) | < 0.001 | 0.0037 (Very small) |
| Respiratory rate | < 0.001 | 0.0234 (Small) |
| Urine output (ml/kg/h) | < 0.001 | 0.0214 (Small) |
| SaFi ratio | < 0.001 | 0.0453 (Small) |
| Modified shock index | < 0.001 | 0.0635 (Moderate) |
| Lactate | < 0.001 | 0.0439 (Small) |
| Hematocrit | < 0.001 | 0.0110 (Small) |
| Bicarbonate | < 0.001 | 0.0311 (Small) |
| Blood pH | < 0.001 | 0.0517 (Small) |
| WBC Count | < 0.001 | 0.0152 (Small) |
| Anion gap | < 0.001 | 0.0148 (Small) |
| Hemoglobin | < 0.001 | 0.0127 (Small) |
| Creatinine | < 0.001 | 0.0409 (Small) |
| Blood urea nitrogen | < 0.001 | 0.0455 (Small) |
| Age | < 0.001 | 0.0049 (Very small) |
| Gender | < 0.001 | Cramer's V = 0.0185 (Small) |

**Table 3.** Percentage of missing values for clinical and laboratory variables across shock subtypes, before imputation. Missingness was assessed after aggregation of measurements into mean values from the first 6 hours of ICU admission. SaFi ratio: ($SpO_2$ / $FiO_2$); WBC: white blood cells; BUN: blood urea nitrogen.

| Variable | No shock | Cardiogenic shock | Hypovolemic shock | Septic shock |
|---|---|---|---|---|
| **Clinical variables** | | | | |
| Temperature | 2.03% | 13.76% | 3.99% | 1.72% |
| Heart rate | 0.81% | 1.82% | 1.48% | 0.68% |
| Systolic blood pressure | 2.47% | 2.49% | 1.82% | 1.10% |
| Diastolic blood pressure | 1.47% | 2.49% | 1.94% | 1.10% |
| Mean arterial pressure | 1.56% | 2.32% | 1.94% | 1.10% |
| Oxygen saturation ($SpO_2$) | 0.91% | 2.99% | 1.71% | 0.99% |
| Respiratory rate | 1.28% | 1.99% | 1.71% | 0.73% |
| Urine output (ml/kg/h) | 43.19% | 24.38% | 26.88% | 19.47% |
| SaFi ratio | 70.42% | 41.63% | 56.49% | 47.49% |
| Modified shock index | 1.58% | 2.32% | 1.94% | 1.20% |
| Age | 0.00% | 0.00% | 0.00% | 0.00% |
| Gender | 0.00% | 0.00% | 0.00% | 0.00% |
| **Laboratory variables** | | | | |
| Lactate | 61.17% | 25.70% | 40.32% | 22.39% |
| Hematocrit | 25.95% | 11.44% | 14.58% | 16.91% |
| Bicarbonate | 25.83% | 8.79% | 17.08% | 13.26% |
| Blood pH | 61.32% | 25.54% | 45.67% | 31.63% |
| WBC count | 28.17% | 11.94% | 19.13% | 18.27% |
| Anion gap | 26.15% | 9.95% | 17.77% | 13.41% |
| Hemoglobin | 28.03% | 11.77% | 18.68% | 18.22% |
| Creatinine | 25.37% | 9.12% | 16.63% | 13.31% |
| BUN | 25.43% | 9.12% | 16.74% | 13.31% |

**Table 4.** Performance metrics (mean ± SD) for each classification model under three imputation strategies (median, k-NN with k=7, and MICE). Pre.: precision; Sen.: sensitivity; F1: F1-score; Spe.: specificity. RF: Random Forest; RL: Logistic Regression; MLP: Multi-Layer Perceptron; XGB: XGBoost.

| Model | Median | | | | k-NN (k = 7) | | | | MICE | | | |
|---|---|---|---|---|---|---|---|---|---|---|---|---|
| | **Pre.** | **Sen.** | **F1** | **Spe.** | **Pre.** | **Sen.** | **F1** | **Spe.** | **Pre.** | **Sen.** | **F1** | **Spe.** |
| ***RF*** | 0.49 ± 0.02 | 0.50 ± 0.02 | 0.49 ± 0.02 | 0.83 ± 0.01 | 0.47 ± 0.02 | 0.48 ± 0.02 | 0.47 ± 0.02 | 0.83 ± 0.01 | 0.50 ± 0.03 | 0.50 ± 0.03 | 0.50 ± 0.03 | 0.83 ± 0.01 |
| **RL** | 0.47 ± 0.02 | 0.48 ± 0.02 | 0.47 ± 0.02 | 0.83 ± 0.01 | 0.46 ± 0.02 | 0.47 ± 0.02 | 0.46 ± 0.02 | 0.82 ± 0.01 | 0.47 ± 0.03 | 0.48 ± 0.02 | 0.47 ± 0.02 | 0.83 ± 0.01 |
| **MLP (64:32)** | 0.41 ± 0.04 | 0.41 ± 0.03 | 0.41 ± 0.04 | 0.80 ± 0.01 | 0.42 ± 0.03 | 0.41 ± 0.02 | 0.41 ± 0.02 | 0.80 ± 0.01 | 0.42 ± 0.03 | 0.42 ± 0.03 | 0.42 ± 0.03 | 0.81 ± 0.01 |
| ***XGB*** | 0.49 ± 0.02 | 0.50 ± 0.02 | 0.49 ± 0.02 | 0.83 ± 0.01 | 0.49 ± 0.02 | 0.50 ± 0.02 | 0.49 ± 0.02 | 0.83 ± 0.01 | 0.47 ± 0.03 | 0.48 ± 0.02 | 0.47 ± 0.03 | 0.83 ± 0.01 |

**Table 5.** Per-class and macro-averaged classification performance for each model (after hyperparameter optimisation) on the test set. AUROC: area under the receiver operating characteristic curve; TPR: true positive rate; TNR: true negative rate; PPV: positive predictive value; NPV: negative predictive value; RF: Random Forest; RL: Logistic Regression; MLP: Multi-Layer Perceptron; XGB: XGBoost; *MLP architecture: 128:64 hidden layers.

| Metric | Model | | | |
|---|---|---|---|---|
| | **RF** | **RL** | **MLP*** | **XGB** |
| AUROC (test Model 1) | 0.83 | 0.83 | 0.83 | 0.82 |
| **No shock identification** | | | | |
| TPR (Sensitivity) | 0.78 | 0.76 | 0.75 | 0.79 |
| TNR (Specificity) | 0.75 | 0.74 | 0.75 | 0.74 |
| PPV (Precision) | 0.86 | 0.86 | 0.86 | 0.86 |
| NPV | 0.62 | 0.60 | 0.62 | 0.64 |
| F1-score | 0.82 | 0.80 | 0.80 | 0.82 |
| **Hypovolemic shock identification** | | | | |
| TPR (Sensitivity) | 0.34 | 0.21 | 0.30 | 0.27 |
| TNR (Specificity) | 0.95 | 0.92 | 0.93 | 0.96 |
| PPV (Precision) | 0.32 | 0.15 | 0.23 | 0.28 |
| NPV | 0.96 | 0.95 | 0.95 | 0.95 |
| F1-score | 0.33 | 0.17 | 0.26 | 0.28 |
| **Cardiogenic shock identification** | | | | |
| TPR (Sensitivity) | 0.57 | 0.45 | 0.53 | 0.60 |
| TNR (Specificity) | 0.97 | 0.90 | 0.95 | 0.97 |
| PPV (Precision) | 0.54 | 0.21 | 0.41 | 0.58 |
| NPV | 0.97 | 0.96 | 0.97 | 0.98 |
| F1-score | 0.56 | 0.29 | 0.47 | 0.59 |
| **Septic shock identification** | | | | |
| TPR (Sensitivity) | 0.76 | 0.51 | 0.69 | 0.76 |
| TNR (Specificity) | 0.86 | 0.88 | 0.86 | 0.87 |
| PPV (Precision) | 0.59 | 0.54 | 0.57 | 0.60 |
| NPV | 0.93 | 0.87 | 0.92 | 0.93 |
| F1-score | 0.66 | 0.52 | 0.62 | 0.67 |
| **Macro-average** | | | | |
| TPR (Sensitivity) | 0.61 | 0.48 | 0.57 | 0.61 |
| TNR (Specificity) | 0.88 | 0.86 | 0.87 | 0.89 |
| PPV (Precision) | 0.58 | 0.44 | 0.52 | 0.58 |
| NPV | 0.87 | 0.85 | 0.87 | 0.88 |
| F1-score | 0.59 | 0.45 | 0.54 | 0.59 |

**Table 6.** Top 5 most important features for each classification model and shock subtype, ranked according to feature importance (Random Forest, MLP, XGBoost and Logistic Regression). BP: blood pressure; BUN: blood urea nitrogen; WBC: white blood cell.

| Rank | Shock vs. No Shock | Septic vs. Other | Hypovolemic vs. Other | Cardiogenic vs. Other |
|---|---|---|---|---|
| **Model: Random Forest** | | | | |
| **1** | Systolic BP | Modified Shock Index | Creatinine | Hematocrit |
| **2** | Modified Shock Index | Temperature | Anion gap | Temperature |
| **3** | Mean arterial pressure | Respiratory rate | Respiratory rate | Hemoglobin |
| **4** | BUN | WBC count | Modified Shock Index | Anion gap |
| **5** | Creatinine | Bicarbonate | Systolic BP | Creatinine |
| **Model: Logistic Regression** | | | | |
| **1** | Modified Shock Index | Mean arterial pressure | Mean arterial pressure | Mean arterial pressure |
| **2** | Systolic BP | Modified Shock Index | Modified Shock Index | Systolic BP |
| **3** | Mean arterial pressure | Diastolic BP | Heart rate | Diastolic BP |
| **4** | Heart rate | Hemoglobin | Hematocrit | Heart rate |
| **5** | BUN | Bicarbonate | Systolic BP | Anion gap |
| **Model: MLP (Neural Network 128:64)** | | | | |
| **1** | Systolic BP | Mean arterial pressure | Hematocrit | Mean arterial pressure |
| **2** | BUN | Heart rate | Hemoglobin | Respiratory rate |
| **3** | Modified Shock Index | Diastolic BP | Creatinine | Systolic BP |
| **4** | Respiratory rate | WBC count | Systolic BP | Hematocrit |
| **5** | Bicarbonate | Hemoglobin | Respiratory rate | Anion gap |
| **Model: XGBoost** | | | | |
| **1** | Systolic BP | WBC count | Systolic BP | Hematocrit |
| **2** | Modified Shock Index | Temperature | Creatinine | Systolic BP |
| **3** | BUN | Modified Shock Index | Respiratory rate | Anion gap |
| **4** | Respiratory rate | Bicarbonate | Anion gap | Mean arterial pressure |
| **5** | Bicarbonate | Mean arterial pressure | Age | Age |

**Table 7**. Row-normalised confusion matrix of the Random Forest cascade model. Rows represent the actual shock classes, and columns represent the predicted shock classes.

| Actual class (n) | No shock | Cardiogenic | Hypovolemic | Septic |
|---|---|---|---|---|
| **No shock (1541)** | 77.5% | 2.9% | 5.1% | 14.5% |
| **Cardiogenic (133)** | 28.6% | 57.1% | 3.8% | 10.5% |
| **Hypovolemic (141)** | 46.1% | 3.5% | 34.0% | 16.3% |
| **Septic (485)** | 17.5% | 3.1% | 3.3% | 76.1% |

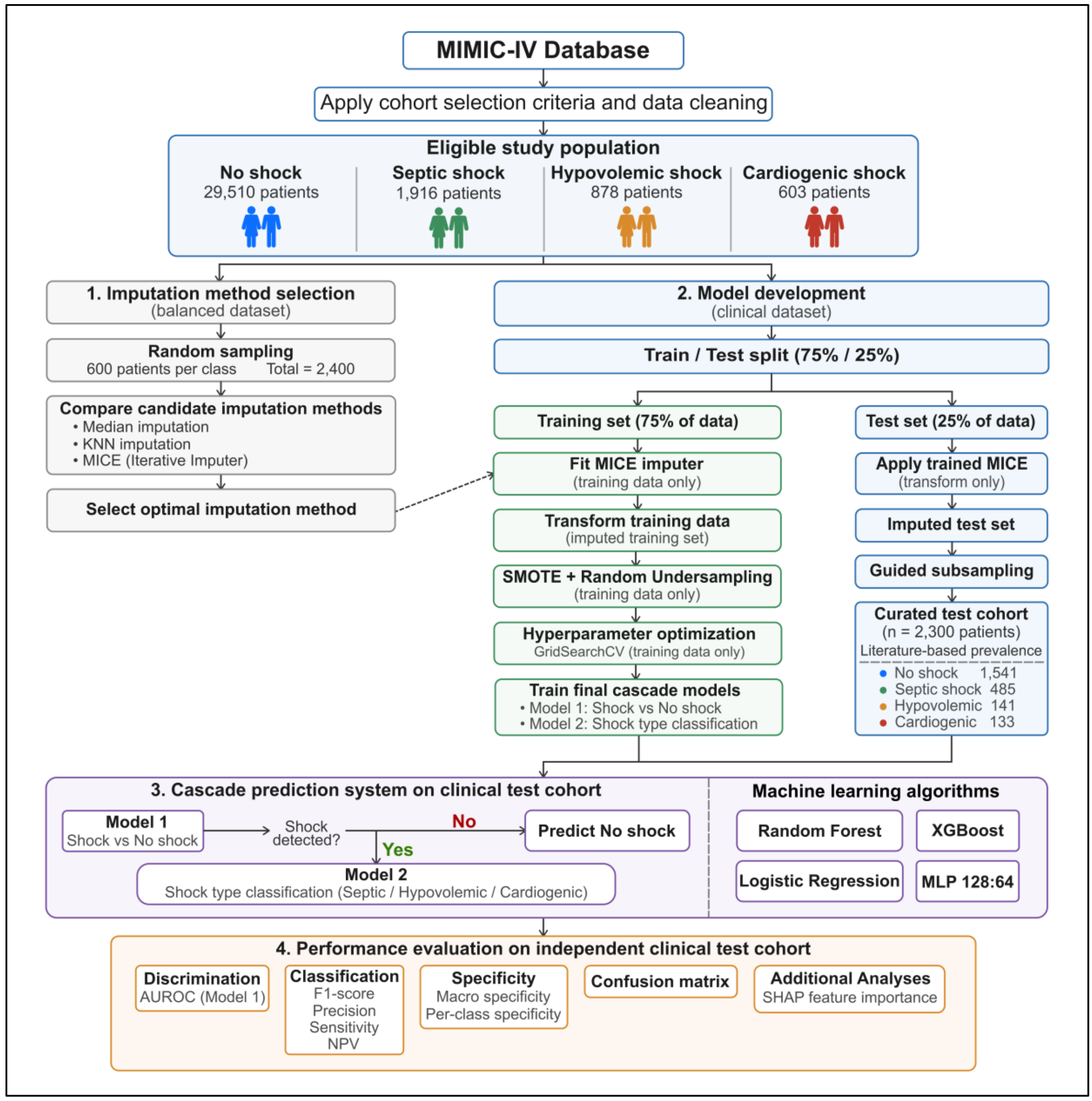

**Figure 1.** Overview of the study pipeline, from cohort selection to model evaluation. MICE: Multiple Imputation by Chained Equations; SMOTE: Synthetic Minority Over-sampling Technique; AUROC: area under the receiver operating characteristic curve; SHAP: SHapley Additive exPlanations.

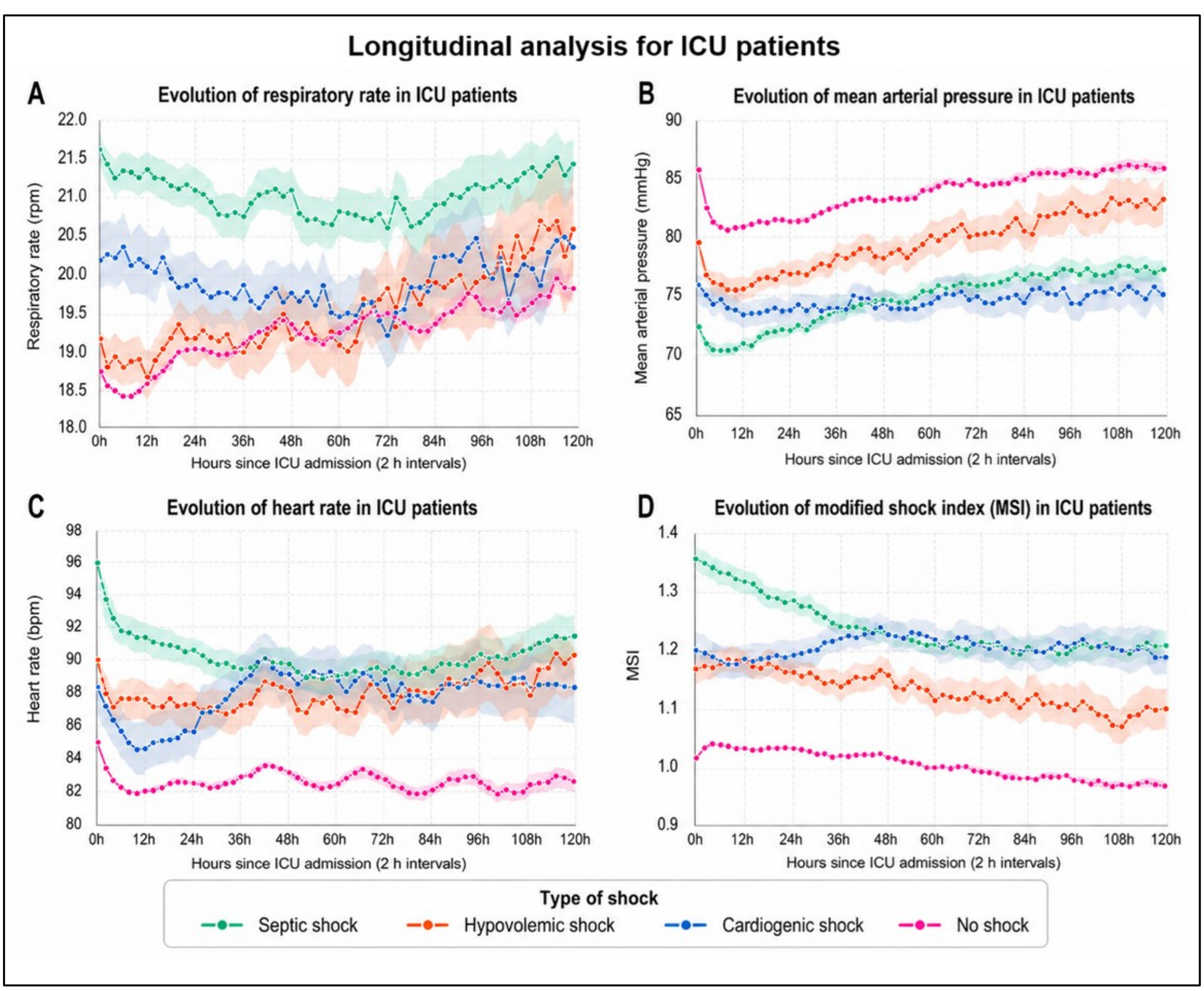

**Figure 2.** Longitudinal evolution of physiological variables during the first 120 hours of ICU admission, stratified by shock subtype. (A) Respiratory rate, (B) mean arterial pressure, (C) heart rate, and (D) modified shock index (MSI). Data are shown as mean values per 2-hour interval, with shaded areas representing the 95% confidence interval.

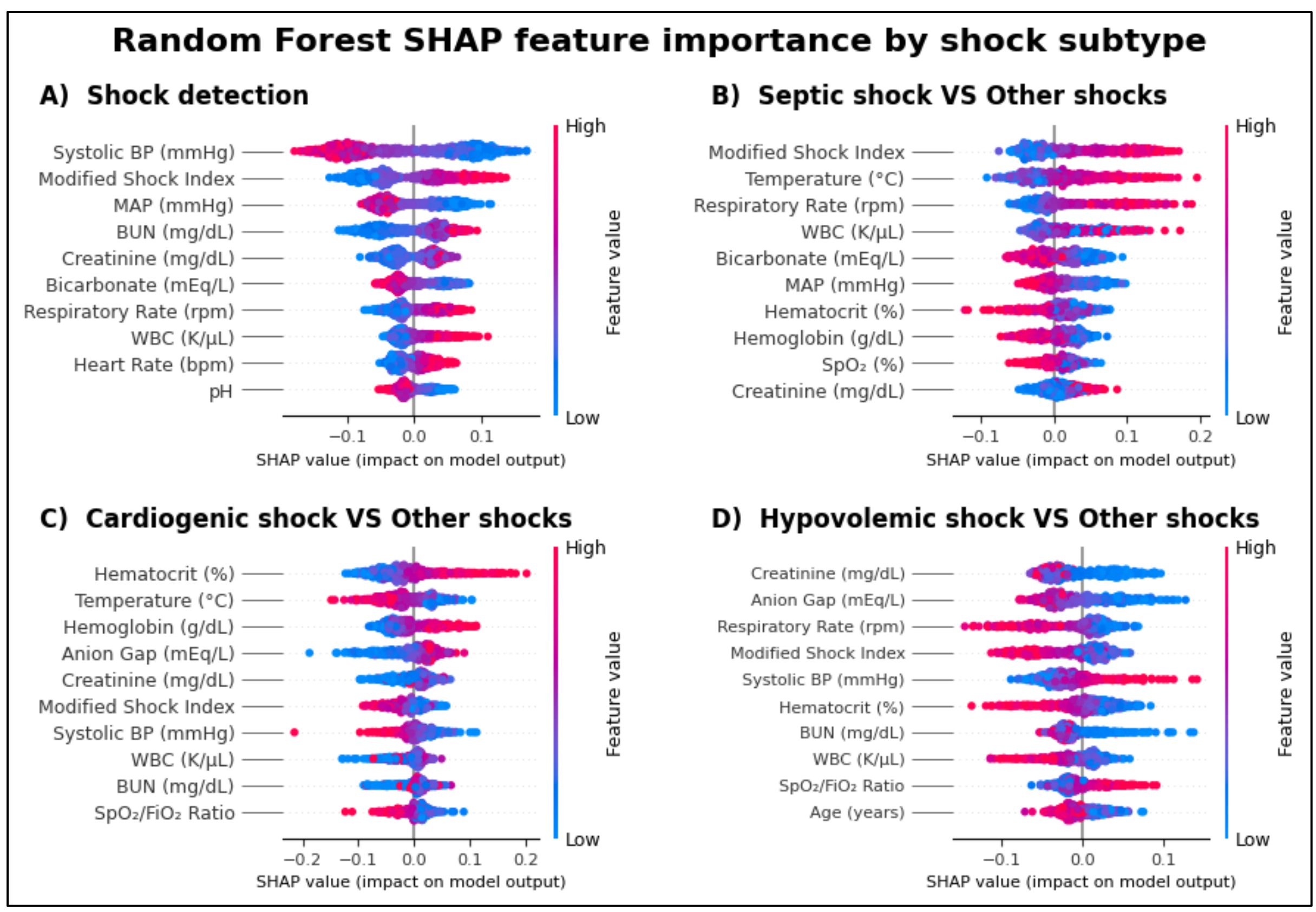

**Figure 3.** SHAP feature importance for the Random Forest model across shock classification tasks. (A) Top 10 features for general shock detection (shock vs. no shock). (B-D) Top 10 features for one-vs-rest classification of (B) septic, (C) cardiogenic, and (D) hypovolemic shock against all other classes. Each point represents a SHAP value for an individual instance, with colour indicating the feature value (red: high, blue: low) and horizontal position indicating its impact on the model output. BP: blood pressure; MAP: mean arterial pressure; BUN: blood urea nitrogen; WBC: white blood cell count; $SpO_2$: oxygen saturation; $FiO_2$: fraction of inspired oxygen; SaFi: $SpO_2/FiO_2$ ratio.

# Appendix A. Supplementary data

**Supplementary Table 1.** Item identifiers (Itemid) from the MIMIC-IV database and plausible value ranges used for outlier filtering of physiological and laboratory variables. NIBP: non-invasive blood pressure; BUN: blood urea nitrogen.

| Variable | Item ID | Plausible value |
|---|---|---|
| **Temperature** | 223761 (oral) and 223762 (rectal) | 30-50 °C |
| **Heart rate** | 220045 | 0-400 bpm |
| **Systolic blood pressure** | 220050 (invasive) and 220179 (NIBP) | 0-300 mmHg |
| **Diastolic blood pressure** | 220051 (invasive) and 220180 (NIBP) | 5-200 mmHg |
| **Mean blood pressure** | 220052 (invasive) and 220181 (NIBP) | 0-200 mmHg |
| **Oxygen saturation** | 220277 | 50-100% |
| **Respiratory rate** | 220210 | 0-50 bpm |
| **Urine output** | 226559 | 0-1500 mL |
| **Lactate** | 50813 | 0-20 mmol/L |
| **Hematocrit** | 51221 | 5-70% |
| **Bicarbonate** | 50882 | 0-50 mmol/L |
| **Blood pH** | 50820 | 6.5-8.0 |
| **White blood cells** | 51301 | 0.1-100 K/µL |
| **Anion gap** | 50868 | 0-80 mmol/L |
| **Hemoglobin** | 51222 | 3-25 g/dL |
| **Creatinine** | 50912 | 0.1-20 mg/dL |
| **BUN** | 51006 | 0-200 mg/dL |

**Heatmap of effect size r / φ for clinical and demographic variables**

| | S vs H | S vs C | H vs C | S vs NS | H vs NS | C vs NS |
|---|---|---|---|---|---|---|
| Systolic BP | 0.228* | 0.008 | 0.241* | 0.234* | 0.083* | 0.134* |
| Modified shock index | 0.275* | 0.208* | 0.035 | 0.237* | 0.087* | 0.077* |
| Mean arterial pressure | 0.227* | 0.137* | 0.089* | 0.221* | 0.076* | 0.090* |
| SaFi index | 0.154* | 0.058* | 0.243* | 0.179* | 0.056* | 0.139* |
| Respiratory rate | 0.244* | 0.111* | 0.135* | 0.149* | 0.019* | 0.053* |
| Diastolic BP | 0.167* | 0.089* | 0.071* | 0.172* | 0.060* | 0.069* |
| Temperature | 0.107* | 0.198* | 0.145* | 0.015* | 0.034* | 0.066* |
| Urine output (ml/kg/h) | 0.168* | 0.074* | 0.081* | 0.144* | 0.029* | 0.049* |
| Heart rate | 0.146* | 0.153* | 0.028 | 0.131* | 0.043* | 0.028* |
| O2 saturation | 0.184* | 0.064* | 0.110* | 0.052* | 0.033* | 0.006 |
| Age | 0.055* | 0.050* | 0.113* | 0.054* | 0.017* | 0.048* |
| Gender (φ) | 0.006 | 0.046 | 0.061* | 0.004 | 0.001 | 0.019* |

Colour scale: 0.05, 0.10, 0.15, 0.20, 0.25, 0.30

S = Septic shock | H = Hypovolemic shock | C = Cardiogenic shock | NS = No shock
(*) = Adjusted p-value (Holm) < 0.05 (significant)

**Supplementary Figure 1.** Heatmap of effect sizes (rank-biserial correlation, r, for continuous variables; phi coefficient, φ, for gender) for pairwise comparisons between shock subtypes and no shock for clinical and demographic variables. S: septic shock; H: hypovolemic shock; C: cardiogenic shock; NS: no shock. Asterisks (*) denote statistically significant differences after Holm correction for multiple comparisons (adjusted $p < 0.05$).

**Heatmap of effect size r for laboratory variables**

| | S vs H | S vs C | H vs C | S vs NS | H vs NS | C vs NS |
|---|---|---|---|---|---|---|
| Creatinine | 0.225* | 0.005 | 0.277* | 0.174* | 0.036* | 0.121* |
| BUN | 0.205* | 0.004 | 0.239* | 0.184* | 0.051* | 0.122* |
| Anion gap | 0.176* | 0.107* | 0.310* | 0.088* | 0.013 | 0.093* |
| Lactate | 0.082* | 0.091* | 0.204* | 0.169* | 0.073* | 0.143* |
| Hematocrit | 0.026 | 0.240* | 0.295* | 0.078* | 0.064* | 0.041* |
| Hemoglobin | 0.025 | 0.243* | 0.245* | 0.098* | 0.058* | 0.028* |
| Bicarbonate | 0.174* | 0.081* | 0.097* | 0.165* | 0.047* | 0.072* |
| Blood pH | 0.096* | 0.005 | 0.104* | 0.198* | 0.089* | 0.122* |
| White blood cells | 0.134* | 0.004 | 0.169* | 0.103* | 0.026* | 0.076* |

0.35 0.30 0.25 0.20 0.15 0.10 0.05

S = Septic shock | H = Hypovolemic shock | C = Cardiogenic shock | NS = No shock
(*) = Adjusted p-value (Holm) < 0.05 (significant)

**Supplementary Figure 2.** Heatmap of effect sizes (rank-biserial correlation, r) for pairwise comparisons between shock subtypes and no shock for laboratory variables. S: septic shock; H: hypovolemic shock; C: cardiogenic shock; NS: no shock; BUN: blood urea nitrogen. Asterisks (*) denote statistically significant differences after Holm correction for multiple comparisons (adjusted $p < 0.05$).